\documentclass{article} %

\usepackage[utf8]{inputenc} %
\usepackage[T1]{fontenc}    %
\usepackage[hyperfootnotes=false]{hyperref}       %
\usepackage{url}            %
\usepackage{booktabs}       %
\usepackage{amsmath,amsfonts,amsthm}       %
\usepackage{mathtools}      %
\usepackage{nicefrac}       %
\usepackage{microtype}      %
\usepackage{subcaption}
\usepackage{bbm}
\usepackage{multirow}
\usepackage[inline]{enumitem}
\usepackage{diagbox}
\usepackage{pifont}
\usepackage[capitalise]{cleveref}  %
\usepackage{comment}
\usepackage{etoolbox}
\usepackage{xcolor}
\usepackage{graphicx}
\usepackage[font=small]{caption}

\PassOptionsToPackage{hyphens}{url}
\usepackage{hyperref}
\hypersetup{colorlinks=true}

\makeatletter\renewcommand\paragraph{\@startsection{paragraph}{4}{\z@}
  {1em \@plus1ex \@minus.2ex}{-.5em}{\normalfont\normalsize\bfseries}}\makeatother

\newcommand*\samethanks[1][\value{footnote}]{\footnotemark[#1]}

\title{RadFusion: Benchmarking Performance and Fairness for Multimodal Pulmonary Embolism Detection from CT and EHR}
\usepackage{authblk}
\author[1]{Yuyin Zhou\thanks{Equal contribution.}}
\author[1]{Shih-Cheng Huang\samethanks}
\author[1]{Jason Alan Fries}
\author[1]{Alaa Youssef}
\author[2]{Timothy J. Amrhein}
\author[1]{Marcello Chang}
\author[3]{Imon Banerjee}
\author[1]{Daniel Rubin}
\author[1]{Lei Xing}
\author[1]{Nigam Shah}
\author[1]{Matthew P. Lungren}

\affil[1]{Stanford University}
\affil[2]{Duke University}
\affil[3]{Arizona State University}

\date{\vspace{-6ex}}

\begin{document}

\maketitle

\begin{abstract}
Despite the routine use of electronic health record (EHR) data by radiologists to contextualize clinical history and inform image interpretation, the majority of deep learning architectures for medical imaging are \emph{unimodal},~\emph{i.e.}, they only learn features from pixel-level information.  
Recent research revealing how race can be recovered from pixel data alone highlights the potential for serious biases in models which fail to account for demographics and other key patient attributes.
Yet the lack of imaging datasets which capture clinical context, inclusive of demographics and longitudinal medical history, has left multimodal medical imaging underexplored.
To better assess these challenges, we present RadFusion, a multimodal, benchmark dataset of 1794 patients with corresponding EHR data and high-resolution computed tomography (CT) scans labeled for pulmonary embolisms.
We evaluate several representative multimodal fusion models and benchmark their fairness properties across protected subgroups, ~\emph{e.g.}, gender, race/ethnicity, age.
Our results suggest that integrating imaging and EHR data can improve classification performance and robustness without introducing large disparities in the true positive rate between population groups.
\end{abstract}

\section{Introduction}

With recent advances in deep learning, AI systems have received unprecedented attention in processing both medical imaging data, ~\emph{e.g.}, computed tomography (CT) scans, and clinical data found in the electronic health records (EHR).
However, architectures which jointly learn feature representations from images and structured EHR data remain underexplored. 
Clinical data availability during image interpretation is particularly important in radiology, where accurate medical diagnosis on imaging relies significantly on pre-test probability, prior diagnosis, clinical and laboratory data, and prior imaging~\cite{huang2020multimodal}. 
For example, a survey showed that more than 85\% of radiologists consider clinical context as vital for radiological exam interpretation~\cite{leslie2000influence}.

While medical imaging benchmarks such as LiTS~\cite{bilic2019liver}, BTCV~\cite{landman2015miccai}, TCIA-Pancreas~\cite{roth2015deeporgan} are ubiquitous in the community, the number of multimodal medical datasets remains limited.
Existing multimodal datasets largely focus on enriching a single modality, such as including multiple imaging protocols or longitudinal image captures for tracking disease progression ~\cite{menze2014multimodal},
or combining medical images with radiology reports (MIMIC-CXR~\cite{johnson2019mimic}, CheXpert~\cite{irvin2019chexpert}, Chest-Xray8~\cite{wang2017chestx}) for chest X-ray diagnosis.
Few datasets provide longitudinal health record data and imaging, with the notable exception of the UK Biobank Imaging Study which includes multi-organ 2D magnetic resonance (MRI), x-ray, and ultrasound imaging in addition to health record data for 100,000 individuals~\cite{littlejohns2020uk}.
The status quo of not providing demographic and clinical history data as part of imaging datasets is concerning as recent research reveals the degree to which pixel data alone encodes protected attributes like race~\cite{banerjee2021reading}, potentially creating hidden biases in medical imaging models.  

To help address these challenges, we present RadFusion, a large-scale multimodal pulmonary embolism database to advance future research into multimodal fusion strategies for integrating 3D medical imaging data and patient EHR data.
Our dataset contains high-quality CT images and patient EHR data collected for 1837 pulmonary embolism cases.
To ensure that all collected cases are representative and of high-quality, the training, validation and testing sets were selected by stratified random sampling from an original cohort pool of 108,991 studies, followed by a careful removal of studies with wrong protocols, significant artifacts, etc.
The ground truth label for each study was curated by two board certified radiologists and verified by a senior radiologist.

To understand the role of each modality, we benchmark the performance of an imaging-only, EHR-only, and a multimodal fusion model using six different evaluation metrics.
We further examine the robustness and fairness of these three architectures by stratifying the performance on key protected patient
subgroups (\emph{i.e.}, gender, age or race).
Specifically, model fairness has been quantified by analyzing equality of opportunity~\cite{hardt2016equality}---the differences in true positive rate (TPR)
across different patient groups.
Our empirical results suggest that multimodal fusion model not only achieves consistent performance gain compared to imaging-based and EHR-based models, but also demonstrates stronger robustness to different population groups without introducing large TPR disparities. 
This finding suggests that instead of developing more advanced medical image representation models, more research interests should also be drawn to designing better multimodal fusion models (~\emph{e.g.}, integrating EHR data and CT).
To facilitate future research on this direction, we release our RadFusion dataset and benchmark different methods in this work.
To summarize, our contributions are three-fold:
\begin{itemize}[leftmargin=*]
   \setlength\itemsep{0.1em}
   \item We release a large-scale multimodal pulmonary embolism detection dataset, RadFusion, consisting of 1837 CT imaging studies (comprising 600,000+ 2D slices) for 1794 patients and their corresponding EHR summary data.   
   To the best of our knowledge, this is the first public dataset combining 3D medical imaging with longitudinal health information extracted from EHRs. 
   \item We benchmark representative imaging-based, EHR-based and multimodal fusion models on the RadFusion dataset, and provide a rigorous ablation study based on six different evaluation measures for a better understanding of the effect of each modality.
   \item We report TPR disparities among different demographic groups to measure model fairness. Our results suggest that the integration of imaging and EHR modalities is a promising direction for improving model performance and robustness without introducing large TPR gaps.
\end{itemize}

%%%%%%%%%%%%%%%%%%%%%%%%%%%%%%%%%%%%%%%%%%%%%%%%%%%%%%%%%%%%%%%%%%%%
% SECTION
%%%%%%%%%%%%%%%%%%%%%%%%%%%%%%%%%%%%%%%%%%%%%%%%%%%%%%%%%%%%%%%%%%%%
\section{Related Works}

\paragraph{Pulmonary embolism detection.} 
Pulmonary Embolism (PE) is known as a life-threatening medical condition, accounting for almost 300,000 hospitalizations and 180,000 deaths in the United States every year~\cite{horlander2003pulmonary}.
Even though the mortality rate among PE patients is high, studies have shown that prompt recognition and immediate initiation of treatments for PE can significantly decrease morbidity and mortality rates ~\cite{leung2012american, leung2011official}.
Definitive diagnosis of PE is made via computed tomography pulmonary angiography (CTPA), which is interpreted manually by radiologists. 
Unfortunately, patients with PE often experience more than 6 days of delay in diagnosis and a quarter of patients are misdiagnosed during their first visit ~\cite{hendriksen2017clinical, alonso2010delay}. 
The long delay and high misdiagnosis rate is in part due to the rapid increase in utilization CTPA (27-fold in emergency settings). 
Many studies have attempted to automate PE diagnosis and patient triaging to alleviate the burden for radiologists ~\cite{yang2019two, tajbakhsh2015computer, huang2020penet}. 
However, few studies have directly included patient clinical history and demographic information as inputs to their model, even though patient EHR is crucial for accurate interpretation of medical images ~\cite{cohen2007accuracy}. 
\vspace{-0.5em}
\paragraph{Multimodal fusion.} 
Pertinent clinical context is essential for providing accurate diagnostic decisions during medical image interpretations. 
Studies have repeatedly shown that clinical history, vitals and laboratory data are crucial for accurate interpretation of medical images, and a lack of access to patient EHR significantly hinders a radiologist's diagnosis ability. 
Similar to radiologists, medical imaging models that can leverage patient EHR may become more clinically relevant and accurate ~\cite{huang2020fusion}. 
Several studies have attempted to integrate clinical context to overcome the limitations of imaging-only models and observed a boost in performance ~\cite{bhagwat2018modeling, li2019early, kawahara2018seven, yala2019deep, an2018comparison, huang2021gloria}. 
However, most of these studies only rely on a few manually selected clinical features. 
\vspace{-0.5em}
\paragraph{Fairness in healthcare.} 
While machine learning models can be powerful tools to deliver automated clinical decision making, they may also exhibit great performance disparities across protected subgroups, leading to different treatment and care delivery~\cite{chen2019can}. 
Equitable use of healthcare data for training machine learning models and applications of algorithmic fairness have received considerable research attention ~\cite{beam2018big,chen2020ethical,pfohl2021empirical,rajkomar2018scalable,tomavsev2019clinically}.
With the recent advance in machine learning, different types of bias have also been identified, such as gender ~\cite{de2019bias} and racial bias~\cite{chouldechova2017fair}.
In the healthcare domain, many studies have also indicated that health disparities widely exist in different protected attributes such as race, sex and age~\cite{kawachi2005health,obermeyer2019dissecting,chen2019can,walter2019age}.
In the medical imaging domain, Seyyed-Kalantari \emph{et al.} examined the TPR disparities of the state-of-the-art X-ray classifiers among different racial groups, and identified that leveraging multiple sources of data can lead to smaller disparities ~\cite{seyyed2020chexclusion}. 
However, to date, the fairness of multimodal medical models has not been studied yet.
Our study provides the first fairness evaluation of multimodal medical model in the use case of pulmonary embolism detection from CT and EHR.

%%%%%%%%%%%%%%%%%%%%%%%%%%%%%%%%%%%%%%%%%%%%%%%%%%%%%%%%%%%%%%%%%%%%
% SECTION: Dataset
%%%%%%%%%%%%%%%%%%%%%%%%%%%%%%%%%%%%%%%%%%%%%%%%%%%%%%%%%%%%%%%%%%%%
\section{Dataset}

\paragraph{Data Acquisition.} 

We retrieved a total of 108,991 CTPA studies from Stanford University Medical Center (SUMC) with the approval of the Stanford Institutional Review Board (IRB). All studies were conducted between 2000 and 2016 and performed under the pulmonary embolism protocol. 
We applied an NLP model by Banejee ~\emph{et al.} ~\cite{banerjee2018radiology,banerjee2019comparative} on the corresponding radiology reports to generate pseudo labels (positive or negative for PE) for all studies. 
Using the generated labels, we retrieved 2500 1.25 mm axial CT, with approximately equal distribution of positive and negative PE labels, from the local picture archiving and communicating system (PACS) for manual review and labeling.
Two radiologists reviewed each study to remove cases with wrong protocol, significant artifacts, poor imaging quality, and non-diagnostic studies. We provide the data characteristic for the 1837 studies that remained in Table. \ref{tab:dataset}.
In addition to CT images, we created a view of each patient's EHR from the SUMC Epic database within an observational window of 12 months prior to their CT examination date, including demographics, vitals, inpatient/outpatient medications,  ICD codes and lab test results. 
We processed these structured EHR (described in Section \ref{subsection:approach}) and provide a summary of the patient's health record as part of our dataset. 
Note that radiology reports are not included in our dataset due to patient privacy concerns and HIPPA compliance.

\vspace{-0.5em}
\paragraph{Annotation.} 
Ground truth labels for all studies were generated by manual review.
Two board certified radiologists (with 8 and 10 years of experience) separately reviewed each CT scan to diagnose the presence of PE, classify the subtype of PE and annotate the slices that contains PE.
The standard descriptions by Remy-Jardin~\emph{et al.}~\cite{remy1997peripheral} were used to define central positive, segmental positive, subsegmental positive and negative PE, which indicates the location or the arterial branch the PE resides.
We made slight modifications to define subsegmental-only PE as the location of the largest defect at the subsegmental level on a spiral CT, allowing a satisfactory visualization of all pulmonary arteries at the segmental level or higher. 
The two radiologists had high inter-rate reliability, with a Cohen's Kappa Score of 0.959. 
For the few cases with conflicting annotations, labels were determined by the more senior out of the two interpreting radiologists.
We randomly partitioned the studies into train/validation/test split of 80\%/10\%/10\% and made sure that no patients overlap between each split.
We provide the detailed data characteristics and splits in Table~\ref{tab:dataset}.

\begin{table}[]
\centering
\footnotesize
 \resizebox{\linewidth}{!}{
\begin{tabular}{|l|l|l|l|l|l|}
\hline

{ Category}                                                                                         & { Sub-category}                                              & { Overall}                                               & { Train}                                                & { Validation}                                            & { Test}                                                  \\ \hline

                                                                                & { \# of studies}                                                        & { 1837}                                                           & { 1454}                                                           & { 193}                                                            & { 190}                                                            \\ \cline{2-6} 

                                                                                & { \# of patients}                                                       & { 1794}                                                           & { 1414}                                                           & { 190}                                                            & { 190}                                                            \\ \cline{2-6} 

\multirow{-3}{*}{ CTPA exams}                                                      & { \begin{tabular}[c]{@{}l@{}}Median \# of slices \\ (IQR)\end{tabular}} & { \begin{tabular}[c]{@{}l@{}}386 \\ (134)\end{tabular}}           & { \begin{tabular}[c]{@{}l@{}}385 \\ (136)\end{tabular}}           & { \begin{tabular}[c]{@{}l@{}}388 \\ (132)\end{tabular}}           & { \begin{tabular}[c]{@{}l@{}}388 \\ (139)\end{tabular}}           \\ \hline

                                                                                & { \# of negative PE}                                                    & { 1111(60.48\%)}                                                  & { 946 (65.06\%)}                                                  & { 85 (44.04\%)}                                                   & { 80 (42.10\%)}                                                   \\ \cline{2-6} 

                                                                                & { \# of positive PE}                                                    & { 726 (39.52\%)}                                                  & { 508 (34.94\%)}                                                  & { 108 (55.96\%)}                                                  & { 110 (57.89\%)}                                                  \\ \cline{2-6} 

                                                                                & { Central}                                                              & { 257(35.40\%)}                                                   & { 202 (39.76\%)}                                                  & { 27 (25.00\%)}                                                   & { 28 (25.45\%)}                                                   \\ \cline{2-6} 

                                                                                & { Segmental}                                                            & { 387(53.31\%)}                                                   & { 281 (55.31\%)}                                                  & { 52 (48.15\%)}                                                   & { 54 (49.09\%)}                                                   \\ \cline{2-6} 

\multirow{-5}{*}{PE}                                                              & { Subsegmental}                                                         & { 82 (11.29\%)}                                                   & { 25 (4.91\%)}                                                    & { 29 (26.85\%)}                                                   & { 28 (25.45\%)}                                                   \\ \hline

                                                                                & { BMI (mean: std)}                                                      & { 28.37 : 9.65}                                                   & { 28.36 : 10.03}                                                  & { 27.11 : 6.78}                                                   & { 29.60 : 9.22}                                                   \\ \cline{2-6} 

\multirow{-2}{*}{Vitals}                                                          & { Pulse (mean: std)}                                                    & { 81.62 : 14.99}                                                  & { 81.53 : 15.64}                                                  & { 83.05 : 11.86}                                                  & { 80.50 : 13.06}                                                  \\ \hline

                                                                                & { D-dimer test taken}                                                   & { 580 (30.62\%)}                                                  & { 461 (30.90\%)}                                                  & { 58 (28.71\%)}                                                   & { 61 (30.50\%)}                                                   \\ \cline{2-6} 

\multirow{-2}{*}{D-dimer}                                                         & { D-dimer positive}                                                     & { 496 (26.18\%)}                                                  & { 389 (26.07\%)}                                                  & { 51 (25.25\%)}                                                   & { 56 (28.00\%)}                                                   \\ \hline
\end{tabular}
}
\caption{Data characteristics of our RadFusion database. Statistics of the training, validation and test set are listed.}
\label{tab:dataset}
\end{table}
\vspace{-1em}
\paragraph{Data usage.}
Our RadFusion dataset can be used for different research purposes such as 1) building better clinical decision models for detecting pulmonary embolism, a life-threatening condition that represents the third most common cause of cardiovascular-related deaths after myocardial infarction and stroke;
2) developing multimodal fusion models using both CT scans and patient EHR, which is relatively under-explored in the field of medical AI, even while utilizing both modalities is crucial for medical image interpretation in real-world clinical settings. 
Notably, we also release different patient demographics information (\emph{e.g}, race, gender) which enables researchers to study model fairness of different machine learning models.
We split the testing set in to Female and Male (based on gender), White and Others\footnote{Others include Black, Asian, Pacific Islander and other groups.} (based on race) , Age $> 80$ and Age $\leq 80$ (based on age). We chose to dichotomize age into categories of older than 80 years and 80 years or younger because prior studies have determined this cutoff as clinically meaningful ~\cite{jimenez2010simplification, lopez2006venous}. The number of instances in each group can be found in Table~\ref{tab:demo_race_age}.
We believe that the broad multimodal data in RadFusion can enable a wide range of research related to foundation models~\cite{bommasani2021opportunities}, leading to potential benefit on different downstream tasks as well.

To the best of our knowledge, our work is the first to provide a fairness evaluation of multimodal fusion models in the medical imaging domain.
We release all de-identified cases, annotation, the splits and the patient attribute information of each study in~\url{https://stanfordaimi.azurewebsites.net/datasets/3a7548a4-8f65-4ab7-85fa-3d68c9efc1bd}.

\begin{table}[]
\centering
\footnotesize
\begin{tabular}{|c|
>{}c |
>{}c |
>{}c |}
\hline
{ \begin{tabular}[c]{@{}l@{}}Patient \\ attribute\end{tabular}} & { Sub-category} & { \begin{tabular}[c]{@{}c@{}}Including \\ subsegmental-only PE\end{tabular}} & \begin{tabular}[c]{@{}c@{}}Excluding\\ subsegmental-only PE\end{tabular} \\ \hline
                                                                                                                     & { Female}                                        & { 99}                                                                                                         & { 87}                                                                                 \\ \cline{2-4} 
\multirow{-2}{*}{Gender}                                                                                             & { Male}                                          & { 91}                                                                                                         & { 75}                                                                                 \\ \hline
                                                                                                                     & { White}                                         & { 119}                                                                                                        & { 95}                                                                                 \\ \cline{2-4} 
\multirow{-2}{*}{Race}                                                                                               & { Others}                                        & { 71}                                                                                                         & { 67}                                                                                 \\ \hline
                                                                                                                     & { $ > 80$}                               & { 57}                                                                                                         & { 47}                                                                                 \\ \cline{2-4} 
\multirow{-2}{*}{Age}                                                                                                & { $\leq 80$}                                        & { 133}                                                                                                         & { 115}                                                                                 \\ \hline
\end{tabular}
\caption{Demographic statistics on the testing set.}
\label{tab:demo_race_age}
\end{table}

\section{Benchmarks}
\label{section:benchmarks}

\subsection{Approaches}
\label{subsection:approach}
\begin{figure*}[htb!]
\begin{center}
\includegraphics[width=0.7\textwidth]{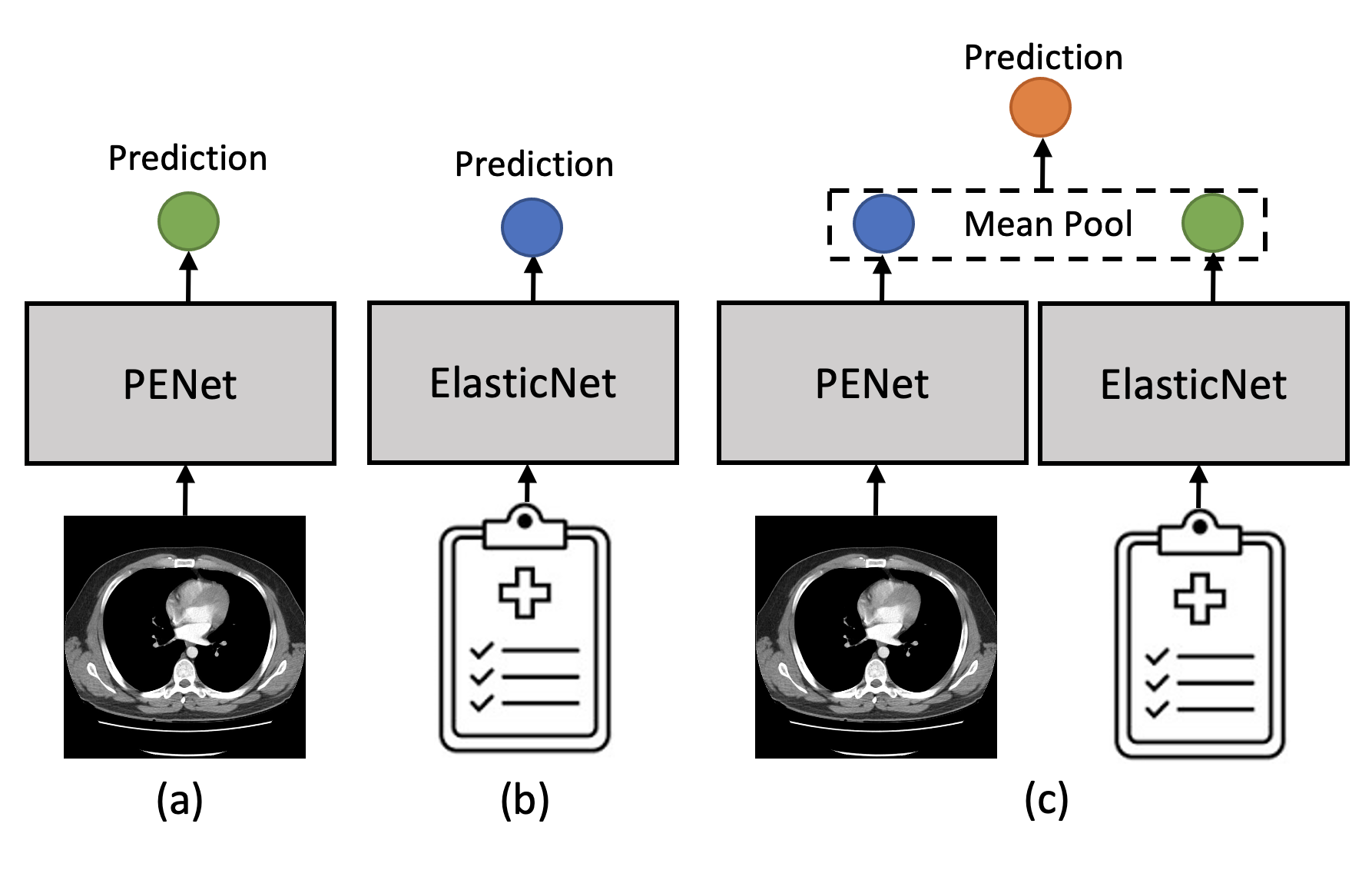}
\end{center}
\vspace{-2em}
  \caption{Different models used for benchmarking: (a) the imaging-only model,  (b) the EHR-only model, and (c) the multimodal fusion model. 
  }
\label{fig:models}
\end{figure*}

To understand the effect of fusion multiple modalities when training medical imaging models, we build 1) an \textbf{imaging-only model} which only consider the pixel values from medical images, 2) an \textbf{EHR-only model} which only relies on patient medical records; and 3) a \textbf{multimodal fusion model} which takes both medical images and EHR data as input for predicting the final outcome.
Figure~\ref{fig:models} illustrates different models used in this study.

\paragraph{Imaging-only model.}

Each CT scan is preprocessed by extracting the pixel data from the original Digital Imaging and Communications in Medicine (DICOM) format and rescaled each slice to $224 \times224$ pixels.
We then apply a viewing window that is optimized for pulmonary arteries (window center = 400, window width = 1000) and clip the Hounsfield Units to the range of $-1000 \sim 900$. 
Lastly, we normalize each CT scan to be zero-centered. 

For our imaging-only model, we use PENet\footnote{Code is publicly available at~\url{https://github.com/marshuang80/penet}}~\cite{huang2020penet}, a 77-layer 3D Convolutional Neural Network (CNN) model capable of detecting PE with high accuracy ~\cite{huang2020penet}. 
PENet is primarily made up of layers of 3D convolutions with skip connections and squeeze-and-excitation blocks. 
Instead of using the entire CT scan, PENet takes in sliding window of CT slices as inputs and made predictions based on the sliding window with the highest prediction probability. 
The detailed model architecture and training procedure can be found in the original manuscript. 
We pretrain PENet with the Kinetics-600 video dataset~\cite{carreira2018short}.

\paragraph{EHR-only model.}

We feature engineered and parsed our EHR data based on the processing steps described by Banerjee~\emph{et al.}~\cite{banerjee2019development}. 
For demographic features, we one-hot encoded gender, race and smoking habits while keeping age as numeric variables. 
All vital features were computed by taking the derivative of the vital values along the temporal axis to represent their sensitivity to change. 
We represent all 641 inpatient and outpatient medications as 1) a binary label of whether the drug was prescribed to the patient and 2) the frequency of prescription within the 12-month window. 
Similarly, ICD codes are also represented as a binary presence/absence label as well as a frequency value. 
For the 22 lab test values, we represent each test with binary presence/absence as well as the latest value of the test. 
We remove ICD codes with less than 1\% occurrences in the training dataset, resulting in 141 diagnosis groups. 
To ensure there are no data leakage, we dropped all ICD codes recorded within 24 hours prior to the CTPA study, in addition to ICD codes recorded with the same encounter number as the patent's CTPA exam. 
All input features are normalized by subtracting the mean and dividing by the standard deviation as a preprocessing step. 
Following ~\cite{huang2020multimodal}, we use an ElasticNet~\cite{zou2005regularization} model that takes in a concatenation of all EHR features as our EHR-only model.

\begin{table}[]
\centering
\footnotesize
\begin{tabular}{|c|c|c|c|}
\hline
 
Metric                    & Imaging-only Model       & EHR-only Model           & Multimodal Fusion Model                        \\
\hline
{ Accuracy}    & { 0.689} & { 0.837} & { \textbf{0.890}}  \\
\hline
{ AUROC}       & { 0.796} & { 0.922} & { \textbf{0.946}}  \\
\hline
{ Specificity} & { 0.863} & { 0.888} & { \textbf{0.900}}  \\
\hline
{ Sensitivity} & { 0.564} & { 0.800} & { \textbf{0.882}}  \\
\hline
{ PPV}         & { 0.849} & { 0.907} & { \textbf{0.924}}        \\
\hline
{ NPV}         & { 0.590} & { 0.763} & { \textbf{0.847}}  \\
\hline
\end{tabular}
\caption{Performance comparison of the imaging-only model, the EHR-only model, and the multimodal fusion model on the testing set. }
\label{tab:multimodal-results}
\end{table}

\vspace{-0.5em}
\paragraph{Multimodal fusion model.}
Features from different modalities can often provide complementary information or separately contribute to the decision making for a machine learning model.
Several different strategies can be leveraged to fuse features from different modalities, including early fusion, late fusion and joint fusion~\cite{huang2020fusion}.
Early fusion combines features from separate modalities at the input level. Joint fusion trains a decision making model using extracted features from single modality models, while propagating the loss back to the feature extracting models.
Late fusion trains separate models for each modality and aggregates the predicted probability from all single-modality models as the final prediction.
In our prior study ~\cite{huang2020multimodal}, we show that late fusion works best for fusing CT and EHR to diagnose PE.
Therefore, for our multimodal fusion design, we take the average of the predicted probabilities from our EHR-only model (ElasticNet) and imaging-only model (PENet) as our final prediction, as shown in Figure~\ref{fig:models}(c).

%%%%%%%%%%%%%%%%%%%%%%%%%%%%%%%%%%%%%%%%%%%%%%%%%%%%%%%%%%%%%%%%%%%%
% SECTION: Results
%%%%%%%%%%%%%%%%%%%%%%%%%%%%%%%%%%%%%%%%%%%%%%%%%%%%%%%%%%%%%%%%%%%%
\subsection{Results and Discussion}
\label{sec:results}

To comprehensively evaluate the performance of the imaging-only model, the EHR-only model and the multimodal fusion model, several evaluation metrics were calculated for the performance across the entire test set, including area under the receiver operating characteristic curve (AUROC), sensitivity, specificity, accuracy, positive predictive value (PPV), and negative predictive value (NPV). We picked the operating point based on the widely adopted Youden's J statistic ~\cite{fluss2005estimation} which maximizes the sum of specificity and sensitivity on our validation set. All 6 evaluation metrics are reported in the following experiments. We use 2 GTX 1080 for all experiments.

\paragraph{Imaging-only vs. EHR-only vs. multimodal fusion.}
To understand the importance of each modality, we test the 3 models on all 190 testing cases.
The results of different models are summarized in Table~\ref{tab:multimodal-results}. 
Out of the two single-modality models, we observe that the EHR-only model consistently achieves better performance on pulmonary embolism detection under all 6 evaluation measures.
When incorporating both the imaging and the EHR modalities, the multimodal fusion model further enhances the performance by a large margin. 
For instance, on the testing set, the imaging-only model and the EHR-only model only yields an accuracy of $68.9\%$ and $83.7\%$ respectively, whereas the multimodal fusion model achieves a much higher accuracy of $89.0\%$.
This suggests that the both modalities are vital for clinical decision making.

\begin{table}[tb!]
\centering
\footnotesize
\begin{tabular}{|c|c|c|c|}
\hline
Metric & { Imaging-only Model} & { EHR-only Model} & Multimodal Fusion Model \\
\hline
Accuracy                                & { 0.759}                            & { 0.877}                        & { \textbf{0.895}}   \\
\hline
AUROC                                   & { 0.842}                            & { 0.932}                        & { \textbf{0.962}}   \\
\hline
Specificity                             & { 0.863}                            & { \textbf{0.888}}               & { 0.838}            \\
\hline
Sensitivity                             & { 0.659}                            & { 0.866}                        & { \textbf{0.951}}   \\
\hline
PPV                                     & { 0.831}                            & { \textbf{0.888}}               & { 0.857}            \\
\hline
NPV                                     & { 0.711}                            & { 0.866}                        & { \textbf{0.944}}  \\
\hline
\end{tabular}
% \vspace{-1em}
\caption{Performance comparison of the imaging-only model, the EHR-only model, and the multimodal fusion model on non-subsegmental-only PE. }
\label{tab:non-segmental-only-PE}
\end{table}
\paragraph{Diagnosis on the non-subsegmental-only PE set.} Diagnosing subsegmental-only PE is known to have questionable clinical significance and is often left untreated~\cite{albrecht2017state}. 
Therefore, we have also computed the same evaluation metrics for the 162 non-subsegmental-only PE cases out of the testing set to understand the clinical utility of our model. 
As shown in Table~\ref{tab:non-segmental-only-PE}, once again, we observe that the EHR-only model outperforms the imaging-only model. But the multimodal fusion model achieves the best results out of the 3 models despite the slightly higher specificity and PPV observed for the EHR-only model.

\begin{table}[tbp!]
\centering
\footnotesize
\begin{tabular}{|c|c|c|c|c|c|c|}
\hline
 
                         & \multicolumn{2}{l|}{Imaging-only Model} & \multicolumn{2}{l|}{EHR-only Model} & \multicolumn{2}{l|}{Multimodal Fusion Model} \\ \cline{2-7} 
 
\multirow{-2}{*}{Metric} & Female                          & Male                          & Female                        & Male                        & Female                             & Male                             \\ \hline
Accuracy                                         & 0.747                           & 0.626                         & 0.869                         & 0.802                       & \underline{\textbf{0.899}}                              & \textbf{0.879}                            \\ \hline
AUROC                                            & 0.844                           & 0.730                          & 0.918                         & 0.924                       & \underline{\textbf{0.953}}                              & \textbf{0.936}                            \\ \hline
Specificity                                      & \underline{\textbf{0.953}}                           & 0.757                         & 0.884                         & 0.892                       & 0.884                              & \textbf{0.919}                            \\ \hline
Sensitivity                                      & 0.589                           & 0.537                         & 0.857                         & 0.741                       & \underline{\textbf{0.911}}                              & \textbf{0.852}                            \\ \hline
PPV                                              & \underline{\textbf{0.943}}                           & 0.763                         & 0.906                         & 0.909                       & 0.911                              & \textbf{0.939}                            \\ \hline
NPV                                              & 0.641                           & 0.528                         & 0.826                         & 0.702                       & \underline{\textbf{0.884}}                              & \textbf{0.810}                             \\ \hline
\end{tabular}
% \vspace{-1em}
\caption{Performance comparison of the imaging-only model, the EHR-only model and the multimodal fusion model for different gender groups on the testing set. \underline{\textbf{Bold underline}} denotes the best results in the \emph{Female} group, \textbf{bold} denotes the best results in the \emph{Male} group.}
\label{tab:perf_comp_gender_all}
\end{table}

\begin{table}[tbp!]
\centering
\footnotesize
\begin{tabular}{|c|c|c|c|c|c|c|}
\hline
 
                         & \multicolumn{2}{c|}{Imaging-only Model} & \multicolumn{2}{c|}{EHR-only Model} & \multicolumn{2}{c|}{Multimodal Fusion Model} \\ \cline{2-7} 
 
\multirow{-2}{*}{Metric} & Female                          & Male                          & Female                        & Male                        & Female                             & Male                             \\ \hline
Accuracy                                         & 0.828                           & 0.680                         & 0.885                         & 0.867                       & \underline{\textbf{0.897}}                              & \textbf{0.893}                            \\ \hline
AUROC                                            & 0.897                           & 0.769                         & 0.934                         & 0.930                       & \underline{\textbf{0.967}}                              & \textbf{0.958}                            \\ \hline
Specificity                                      & \underline{\textbf{0.953}}                           & 0.757                         & 0.884                         & \textbf{0.892}                       & 0.837                              & 0.838                            \\ \hline
Sensitivity                                      & 0.705                           & 0.605                         & 0.886                         & 0.842                       & \underline{\textbf{0.955}}                              & \textbf{0.947}                            \\ \hline
PPV                                              & \underline{\textbf{0.939}}                           & 0.719                         & 0.886                         & \textbf{0.889}                       & 0.857                              & 0.857                            \\ \hline
NPV                                              & 0.759                           & 0.651                         & 0.884                         & 0.846                       & \underline{\textbf{0.947}}                              & \textbf{0.939}                            \\ \hline
\end{tabular}
\caption{Performance comparison of the imaging-only model, the EHR-only model and the multimodal fusion model for different gender groups on the non-subsegmental-only PE set. \underline{\textbf{Bold underline}} denotes the best results in the \emph{Female} group, \textbf{bold} denotes the best results in the \emph{Male} group.}
% \vspace{-1em}
\label{tab:perf_comp_gender}
\end{table}

\section{Fairness Evaluation}
Assessing fairness and robustness of machine learning systems is critical prior to deployment~\cite{seyyed2020chexclusion,kadambi2021achieving,obermeyer2019dissecting}.
Although the multimodal fusion model achieves better performance compared with the imaging-only model and the EHR-only model, it is not clear whether this benefit can be preserved for different population groups with sensitive attributes.
In addition, it remains unknown whether the integration of both modalities will exacerbate existing model biases against different patient attributes in gender, race, \emph{etc.}, which can further raise ethical concerns when deploying such models~\cite{kadambi2021achieving}.

To answer these questions, in this section, we provide a rigorous fairness analysis of the 3 different models for various patient groups for both  the full test set and the non-subsegmental-only test set. 
We first report the performance comparison of different models in AUROC, sensitivity, specificity, accuracy, PPV and NPV for each gender group in Table~\ref{tab:perf_comp_gender_all} and Table~\ref{tab:perf_comp_gender}.
We observe that, under both evaluation settings, the multimodal fusion model achieves better results for different gender groups compared with the imaging-only model and the EHR-only model by a large margin. 
In particular, on the \emph{Male} group, the multimodal fusion model achieves accuracies of $87.9\%$ and $89.3\%$ on the testing set and non-subsegmental-only PE, outperforming the EHR-only model by $7.7\%$ and $2.6\%$ respectively.
Similarly, we demonstrate that the multimodal fusion model achieves the best performance for different age groups in Table~\ref{tab:perf_comp_age_all} \& Table~\ref{tab:perf_comp_age},  and for different racial groups in Table~\ref{tab:perf_comp_race_all} \& Table~\ref{tab:perf_comp_race}.
This fact indicates that the integration of both the imaging modality and the EHR modality not only improves the standard performance but also consistently yields more robust results against different population groups.

\begin{table}[tbp!]
\centering
\footnotesize
\begin{tabular}{|c|c|c|c|c|c|c|}
\hline
 
                         & \multicolumn{2}{l|}{Imaging-only Model} & \multicolumn{2}{l|}{EHR-only Model} & \multicolumn{2}{l|}{Multimodal Fusion Model} \\ \cline{2-7} 
 
\multirow{-2}{*}{Metric} & Age $ > 80$                           & Age $\leq 80$                          & Age $ > 80$                        & Age $\leq 80$                        & Age $> 80$                             & Age $\leq 80$                             \\ \hline
Accuracy                                         & 0.737                           & 0.669                         & 0.825                          & 0.842                       & \underline{\textbf{0.895}}                              & \textbf{0.887}                            \\ \hline
AUROC                                            & 0.886                           & 0.766                          & 0.933                         & 0.917                      & \underline{\textbf{0.979}}                              & \textbf{0.936}                            \\ \hline
Specificity                                      & \underline{\textbf{1.0}}                           & 0.823                         & 0.944                         & 0.871                       & 0.944                              & \textbf{0.887}                        \\ \hline
Sensitivity                                      & 0.615                            & 0.535                         & 0.769                         & 0.817                       & \underline{\textbf{0.872}}                              & \textbf{0.887}                            \\ \hline
PPV                                              & \underline{\textbf{1.0}}                           & 0.776                         & 0.968                         & 0.879                       & 0.971                              & \textbf{0.900}                            \\ \hline
NPV                                              & 0.545                           & 0.607                         & 0.654                         & 0.806                       & \underline{\textbf{0.773}}                              & \textbf{0.873}                             \\ \hline
\end{tabular}
\caption{Performance comparison of the imaging-only model, the EHR-only model and the multimodal fusion model for different age groups on the testing set. \underline{\textbf{Bold underline}} denotes the best results in the \emph{Age} $ > 80$ group, \textbf{bold} denotes the best results in the \emph{Age} $ \leq 80$ group.}
\label{tab:perf_comp_age_all}
\end{table}

\begin{table}[tbp!]
\centering
\footnotesize
\begin{tabular}{|c|c|c|c|c|c|c|}
\hline
 
                         & \multicolumn{2}{c|}{Imaging-only Model} & \multicolumn{2}{c|}{EHR-only Model} & \multicolumn{2}{c|}{Multimodal Fusion Model} \\ \cline{2-7} 
 
\multirow{-2}{*}{Metric} & Age $ > 80$                          & Age $ \leq 80$                          & Age $ > 80$                        & Age $ \leq 80$                        & Age $ > 80$                             & Age $ \leq 80$                             \\ \hline
Accuracy                                         & 0.809                           & 0.739                         & 0.894                         & 0.870                       & \underline{\textbf{0.957}}                              & \textbf{0.870}                            \\ \hline
AUROC                                            & 0.925                           & 0.814                         & 0.935                        & 0.930                       & \underline{\textbf{0.989}}                              & \textbf{0.953}                            \\ \hline
Specificity                                      & \underline{\textbf{1.0}}                           & \textbf{0.823}                         & 0.944                        & 0.871                       & 0.944                              & 0.806                            \\ \hline
Sensitivity                                      & 0.690                           & 0.642                         & 0.862                         & 0.868                       & \underline{\textbf{0.966}}                              & \textbf{0.943}                            \\ \hline
PPV                                              & 1.0                           & 0.756                         & 0.962                        & \textbf{0.852}                       & \underline{\textbf{0.966}}                              & 0.806                            \\ \hline
NPV                                              & 0.667                           & 0.729                        & 0.810                         & 0.885                       & \underline{\textbf{0.944}}                              & \textbf{0.943}                            \\ \hline
\end{tabular}
\caption{Performance comparison of the imaging-only model, the EHR-only model and the multimodal fusion model for different age groups on the non-subsegmental-only PE set. \underline{\textbf{Bold underline}} denotes the best results in the \emph{Age} $ > 80$ group, \textbf{bold} denotes the best results in the \emph{Age} $ \leq 80$ group.}
\label{tab:perf_comp_age}
\end{table}

\begin{table}[htbp]
\centering
\footnotesize
\begin{tabular}{|c|c|c|c|c|c|c|}
\hline
 
                         & \multicolumn{2}{l|}{Imaging-only Model} & \multicolumn{2}{l|}{EHR-only Model} & \multicolumn{2}{l|}{Multimodal Fusion Model} \\ \cline{2-7} 
 
\multirow{-2}{*}{Metric} & White                           & Others                         & White                        & Others                       & White                            & Others                             \\ \hline
Accuracy                                         & 0.622                           & 0.803                         & 0.832                         & 0.845                       & \underline{\textbf{0.882}}                              & \textbf{0.901}                            \\ \hline
AUROC                                            & 0.76                           & 0.852                          & 0.932                         & 0.914                      & \underline{\textbf{0.944}}                              & \textbf{0.959}                            \\ \hline
Specificity                                      & 0.825                           & 0.900                         & 0.900                          & 0.875                       & \underline{\textbf{0.900}}                              & \textbf{0.900}                            \\ \hline
Sensitivity                                      & 0.519                            & 0.677                         & 0.797                         & 0.806                        & \underline{\textbf{0.873}}                              & \textbf{0.903}                            \\ \hline
PPV                                              & 0.854                           & 0.840                         & 0.940                         &  0.833                       & \underline{\textbf{0.945}}                              & \textbf{0.875}                            \\ \hline
NPV                                              & 0.465                          & 0.783                         & 0.692                         & 0.854                       & \underline{\textbf{0.783}}                              & \textbf{0.923}                             \\ \hline
\end{tabular}
\caption{Performance comparison of the imaging-only model, the EHR-only model and the multimodal fusion model for different racial groups on the testing set. \underline{\textbf{Bold underline}} denotes the best results in the \emph{White} group, \textbf{bold} denotes the best results in the \emph{Others} group.}
\label{tab:perf_comp_race_all}
\end{table}

\begin{table}[htbp]
\centering
\footnotesize
\begin{tabular}{|c|c|c|c|c|c|c|}
\hline
 
                         & \multicolumn{2}{c|}{Imaging-only Model} & \multicolumn{2}{c|}{EHR-only Model} & \multicolumn{2}{c|}{Multimodal Fusion Model} \\ \cline{2-7} 
 
\multirow{-2}{*}{Metric} & White                          & Others                          & White                       & Others                        & White                            & Others                             \\ \hline
Accuracy                                         & 0.695                           & 0.851                         & 0.895                         & 0.851                       & \underline{\textbf{0.905}}                              & \textbf{0.881}                            \\ \hline
AUROC                                            & 0.809                          & 0.892                         & 0.944                         & 0.91                       & \underline{\textbf{0.964}}                              & \textbf{0.962}                            \\ \hline
Specificity                                      & 0.825                           & \textbf{0.900}                        & \underline{\textbf{0.900}}                         & 0.875                      & 0.850                              & 0.825                            \\ \hline
Sensitivity                                      & 0.600                           & 0.778                         & 0.891                         & 0.815                       & \underline{\textbf{0.945}}                              & 0.963                            \\ \hline
PPV                                              & 0.825                           & \textbf{0.840}                         & \underline{\textbf{0.925}}                         & 0.815                       & 0.897                              & 0.788                            \\ \hline
NPV                                              & 0.600                           & 0.857                         & 0.857                         & 0.875                       & \underline{\textbf{0.919}}                              & \textbf{0.971}                            \\ \hline
\end{tabular}
\caption{Performance comparison of the imaging-only model, the EHR-only model and the multimodal fusion model for different racial groups on the non-subsegmental-only PE set. \underline{\textbf{Bold underline}} denotes the best results in the \emph{White} group, \textbf{bold} denotes the best results in the \emph{Others} group.}
\label{tab:perf_comp_race}
\end{table}

In addition to overall model performance for different patient groups, we also assess the fairness of different models by reporting the equal opportunity difference (EOD) which measures the difference in TPR (\emph{i.e.}, Sensitivity) for the privileged and under-privileged groups following the evaluation protocol in~\cite{hardt2016equality,seyyed2020chexclusion,li2021estimating}.
We choose to use the TPR gap as our fairness metric based on the needs of the clinical diagnostic setting---high TPR disparity indicates that sick members from one demographic group would not be given correct diagnoses at the same rate as the general population, which can be dangerous for clinical deployment.
The evaluation results on the test set and on non-subsegmental-only PE cases are summarized in Table~\ref{tab:EOD_all} and Table~\ref{tab:EOD} respectively.
They suggest that while the imaging-only model and the EHR-only model can yield large TPR gaps for different gender and racial groups, the multimodal fusion model consistently yields TPR gaps $< 6\%$. For instance, the TPR gaps of different racial groups from the imaging-only model can be as large as $15.8\%$ and $17.8\%$ on the testing set and on non-subsegmental-only cases, and the TPR gap of different gender groups from the EHR-only model reaches $11.6\%$ on the testing set.
As a comparison, the largest TPR gap observed for the multimodal fusion model is only $5.9\%$, for different gender groups on the testing set.
However, we note that the multimodality fusion model does not consistently improve the fairness compared with single-modality models. How to design fairer multimodality fusion models remain as an open question.

\paragraph{Discussion.} Although our result
demonstrates the huge potential for integrating both the imaging and the EHR modalities to improve model performance and robustness without introducing large TPR disparities between different population groups, several gaps are yet to be addressed. 
Future studies should examine fusion model
designs that not only improve overall performance but also reduce model bias when integrating different modalities. Moreover, given the inherent biases in existent large public datasets, it would be important to investigate the performance of multimodal fusion models on diverse multi-source datasets in an effort to improve algorithm fairness for the different gender, racial and age groups.

\paragraph{Limitation.} One limitation of our work is that the different population groups collected in the dataset are not well-balanced, which may have contributed to the evaluation of the TPR gaps. Additionally, future research should also extend the dataset to external institutions with new scanners and protocols.

\begin{table}[tbp!]
\centering
\footnotesize
\begin{tabular}{|c|c|c|c|c|c|c|}
\hline
 
Patient Attribute & \multicolumn{2}{c|}{Imaging-only Model} & \multicolumn{2}{c|}{EHR-only Model} & \multicolumn{2}{c|}{Multimodal Fusion Model} \\ \hline
Gender            & \multicolumn{2}{c|}{\textbf{0.052}}                                      & \multicolumn{2}{c|}{0.116}                                  & \multicolumn{2}{c|}{0.059}                                            \\ \hline
Race              & \multicolumn{2}{c|}{0.158}                                      & \multicolumn{2}{c|}{\textbf{0.009}}                                  & \multicolumn{2}{c|}{0.030}                                            \\ \hline
Age               & \multicolumn{2}{c|}{0.080}                                      & \multicolumn{2}{c|}{0.048}                                  & \multicolumn{2}{c|}{\textbf{0.015}}                                            \\ \hline
\end{tabular}
\vspace{-.8em}
\caption{EOD under various patient attributes (\emph{i.e.}, gender, race, age) on the testing set. Less EOD indicates less bias. }
\label{tab:EOD_all}
\end{table}

\begin{table}[tbp!]
\centering
\footnotesize
\begin{tabular}{|c|c|c|c|c|c|c|}
\hline
 
Patient Attribute & \multicolumn{2}{c|}{Imaging-only Model} & \multicolumn{2}{c|}{EHR-only Model} & \multicolumn{2}{c|}{Multimodal Fusion Model} \\ \hline
Gender            & \multicolumn{2}{c|}{0.100}                                      & \multicolumn{2}{c|}{0.044}                                  & \multicolumn{2}{c|}{\textbf{0.008}}                                            \\ \hline
Race              & \multicolumn{2}{c|}{0.178}                                      & \multicolumn{2}{c|}{0.076}                                  & \multicolumn{2}{c|}{\textbf{0.017}}                                            \\ \hline
Age              & \multicolumn{2}{c|}{0.047}                                      & \multicolumn{2}{c|}{\textbf{0.006}}                                  & \multicolumn{2}{c|}{0.023}                                            \\ \hline
\end{tabular}
\vspace{-.8em}
\caption{EOD under various patient attributes (\emph{i.e.}, gender, race, age) on the non-subsegmental-only PE set. Less EOD indicates less bias.}
\label{tab:EOD}
\end{table}

\section{Conclusion}
Despite the pervasiveness of medical imaging benchmarks, the number of multimodal medical datasets remains limited. 
Datasets involving both medical imaging and EHR patient data are even more scarce.
To advance research in this field, in this study, we introduce \textbf{RadFusion}, a large-scale multimodal dataset consisting of paired CT images and EHR patient data from over 1800 studies. 
We identify the importance of both imaging data and EHR patient data for pulmonary embolism detection by benchmarking different representative imaging-based, EHR-based and multimodal fusion models on RadFusion.
Through extensive evaluation on both the testing set and non-subsegmental-only PE cases, our initial results suggest that compared with single-modality methods, the multimodal fusion model can significant improve performance and robustness, without introducing large TPR disparities between different population groups.

To the best of our knowledge, RadFusion is the first public database focusing on combining medical imaging data with large-scale patient EHR for advancing clinical diagnosis.
To support better clinical utility, RadFusion also provides opportunities to consider fairness and robustness against different patient attributes when designing multimodal fusion models.
We hope our study can serve as an important baseline and facilitate future research on this direction.

\paragraph{Acknowledgement}
Research reported in this publication was supported by the National Heart, Lung, And Blood Institute of the National Institutes of Health under Award Number R01HL155410 and the National Library Of Medicine of the National Institutes of Health under Award Number R01LM012966. The content is solely the responsibility of the authors and does not necessarily represent the official views of the National Institutes of Health.

\bibliographystyle{plain}
{\small
	\bibliography{egbib}
}

\end{document}